\documentclass[]{aastex631}

\usepackage{verbatim}
\usepackage{amsmath}
\usepackage{graphicx}
\usepackage{url}
\usepackage{footmisc}

\begin{document}

\title{False Alarms Revealed in a Planet Search of TESS Light Curves}

\correspondingauthor{Michelle Kunimoto}
\email{mkuni@mit.edu}

\author[0000-0001-9269-8060]{Michelle Kunimoto}
\affiliation{Kavli Institute for Astrophysics and Space Research, Massachusetts Institute of Technology, Cambridge, MA 02139, USA}

\author[0000-0003-0081-1797]{Steve Bryson}
\affiliation{NASA Ames Research Center, Moffett Field, CA 94035, USA}

\author[0000-0002-6939-9211]{Tansu Daylan}
\altaffiliation{LSSTC Catalyst Fellow}
\affiliation{Department of Astrophysical Sciences, Princeton University, 4 Ivy Lane, Princeton, NJ 08544}

\author[0000-0001-6513-1659]{Jack J. Lissauer}
\affiliation{NASA Ames Research Center, Moffett Field, CA 94035, USA}

\author[0000-0002-1119-7473]{Michael R. B. Matesic}
\affiliation{Department of Physics and Astronomy, Bishop's University, 2600 Rue College, Sherbrooke, QC J1M 1Z7, Canada}

\author[0000-0001-7106-4683]{Susan E. Mullally}
\affiliation{Space Telescope Science Institute, 3700 San Martin Drive, Baltimore, MD, 21218, USA}

\author[0000-0002-5904-1865]{Jason F. Rowe}
\affiliation{Department of Physics and Astronomy, Bishop's University, 2600 Rue College, Sherbrooke, QC J1M 1Z7, Canada}

\begin{abstract}

We examined the period distribution of transit-like signatures uncovered in a Box-Least Squares transit search of TESS light curves, and show significant pileups at periods related to instrumental and astrophysical noise sources. Signatures uncovered in a search of inverted light curves feature similar structures in the period distribution. Automated vetting methods will need to remove these excess detections, and light curve inversion appears to be a suitable method for simulating false alarms and designing new vetting metrics.
\end{abstract}


\section{Introduction}\label{sec:intro}
Exoplanet demographics provide significant insight into exoplanet formation and evolution theory.  Exoplanet catalogs, however, do not represent the true exoplanet population.  Catalogs are not complete because they miss exoplanets, and may not be reliable because of contamination from false positives (e.g., nearby or bound eclipsing binaries) or false alarms (e.g., noise, systematics, and stellar variability).  The Kepler mission \citep{Borucki2010} was designed to produce demographic studies, yielding a catalog with well-measured completeness and reliability \citep{Thompson2018}, making it possible to infer the underlying planet population.  The TESS mission \citep{Ricker2015} is finding thousands of exoplanets, but the TESS Object of Interest catalog is not useful for demographics because it has unknown completeness and reliability.  

Exoplanet catalogs are typically produced by a software pipeline that detects transit events which are vetted to determine if the events are caused by planets.  The most straightforward way to produce a catalog with well-measured completeness and reliability is to use a fully automated detection and vetting pipeline.  Designing an automated vetting pipeline is a non-trivial task requiring an understanding of false alarms found in observed data. Automated vetting \citep{Thompson2018} becomes particularly challenging in the low signal-to-noise ratio (S/N) regime of smaller and/or longer period planets because most false alarms due to noise and systematics have low S/N.

In this paper, we investigate the impact of false alarms on planet searches of TESS data by performing a search of a large number of stars.  We use light curves obtained from multiple sources to investigate whether or not systematics depend on how the light curves were generated. We identify periods with excess detections, or {\it pileups}, and speculate about their causes in TESS systematics. We also perform a preliminary exploration using light curve inversion to simulate false alarms, needed for characterizing catalog reliability for future TESS demographics studies.

\section{Planet Search}

We perform our investigation on M dwarfs ($R < 0.6~R_{\odot}$, $M < 0.6~M_{\odot}$, $T_{\text{eff}} < 3900~K$), which are of high interest to exoplanet searches, especially for small planets. We selected all 92,899 M dwarfs brighter than $T = 13.5$ mag from the TESS Input Catalog \cite[TICv8.2;][]{Paegert2021} that were observed by TESS across sectors 1 to 52. Multi-sector light curves from the Quick-Look Pipeline \cite[QLP;][]{Huang2020,Kunimoto2021,Kunimoto2022} were obtained for each star. To compare to QLP, light curves were also extracted using the \texttt{eleanor} pipeline \citep{Feinstein2019}. Both \texttt{eleanor} CORR\textunderscore FLUX and PCA\textunderscore FLUX systematics-corrected versions of each light curve were saved.

Long-term trends from all three sets of light curves (QLP, CORR\textunderscore FLUX, and PCA\textunderscore FLUX) were removed using a biweight detrending algorithm with a 0.5-day window length as implemented in \texttt{wotan} \citep{Hippke2019}. We then performed a multi-planet search of the detrended light curves with the Box-Least Squares (BLS) algorithm implemented in \texttt{cuvarbase}\footnote{\url{https://johnh2o2.github.io/cuvarbase/}}. Each star was searched from 0.5 days to 100 days or the length of the longest consecutive stretch of TESS sectors comprising the light curve, whichever was shorter. This strategy was chosen to minimize the challenges of searching data with long gaps between sectors. If a detection had S/N$~>9$, data within one transit duration of the measured center of the transit time is masked out and the residuals were searched, up to five times per light curve. We define Threshold Crossing Events (TCEs) as signals with S/N$~>9$ and at least two events.

The same BLS procedure was re-run on the same light curves after inversion to produce inverted TCEs (invTCEs). Inversion should eliminate valid transits but preserve the general noise and systematic properties seen in the original time series, and was a strategy used by Kepler to simulate false alarms \citep{Thompson2018}.

\begin{figure}
    \centering
    \includegraphics[width=0.98\linewidth]{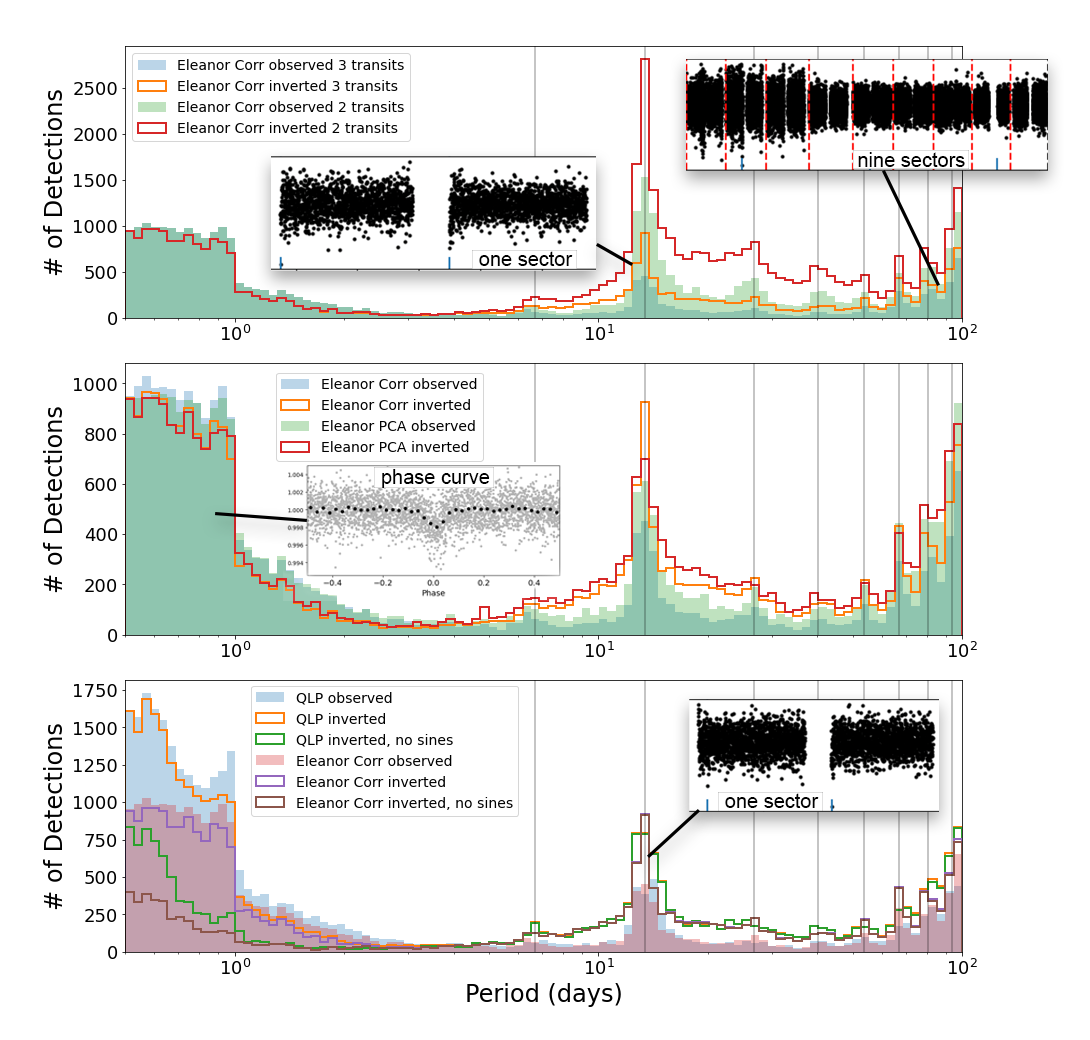}
    \caption{Period histograms of TCEs and invTCEs identified in various input TESS light curves by our detection pipeline, with insets showing inverted \texttt{eleanor} CORR\textunderscore FLUX light curves with planet-like false alarms.  Transits in the insets are indicated by short blue vertical lines, and the red vertical dashed lines delineate sectors. The filled histograms are TCEs and the line histograms are invTCEs.  The vertical gray lines are aliases of the 13.7-day TESS orbital period.  Most signals are false alarms.  We see that, in all cases, pileups in TCEs correspond to pileups in invTCEs, indicating that many detections in the observed data are false alarms.  \textbf{Top panel:} CORR\textunderscore FLUX light curves, comparing two- and three-transit thresholds for transit detection.  \textbf{Middle panel:} Comparing three-transit detections from CORR\textunderscore FLUX and PCA\textunderscore FLUX light curves.  \textbf{Lower panel:} Comparing three-transit detections from CORR\textunderscore FLUX and QLP light curves.  The lower panel also shows ``no sines'' TCEs and invTCEs after the removal of sinusoidal signals identified by the SWEET test (see text).}
    \label{fig:histogram}
\end{figure}

\section{Results}

Figure~\ref{fig:histogram} shows histograms of TCE and invTCE periods from all three light curve sources. In all cases, we see the same general structure of pileups, which are similar to those found by the SPOC pipeline \citep{Jenkins2016} using postage stamp targets\footnote{\url{https://archive.stsci.edu/tess/tess_drn.html}}. While the inverted light curves include detections of astrophysical phenomena such as stellar flares and some eclipsing binaries, detection pileups occur at periods not correlated with flare statistics. We conclude that light curve inversion is a suitable method for simulating TESS false alarms in future demographic studies.

We see two classes of pileups: 
\begin{itemize}
    \item Sharp peaks with broad shoulders at aliases of the 13.7-day TESS orbital period.  Inspection of observed and inverted light curves reveals that most of these detections are easily identified excursions at the edges of data gaps, due to scattered light, that were not removed by detrending.  Not all residual edge effects are easily identified, however, as shown by the inset light curve examples in Figure~\ref{fig:histogram}.  
    \item A general excess with a period $<$ 1 day.  Inspection of the light curves reveals that most of these are highly periodic, which we believe are due primarily to scattered light from the Earth's rotation, and also stellar variability (e.g., spots and pulsations) and contact eclipsing binaries.  We find that the Kepler SWEET test \citep{Thompson2018}, which measures the correlation of light curves with sinusoids, enables the removal of the highly sinusoidal detections, shown by the ``no sines'' histograms in Figure~\ref{fig:histogram}.
\end{itemize}

Comparing the pileup structures for the various light curve sources, we see little difference between \texttt{eleanor} CORR\textunderscore FLUX and PCA\textunderscore FLUX light curves.  For periods $\gtrsim3$ days, we see little difference between \texttt{eleanor} and QLP, but QLP light curves have a larger excess with periods below one day.  We find that requiring three vs. two transits for a transit detection significantly reduces the number of detections with period $\gtrsim3$ days, but does not remove the pileup structure.  While most excess detections below one day are sinusoidal, some appear more transit like, as shown in the inset example in Figure~\ref{fig:histogram}.

\section{Conclusions}

Creating an exoplanet catalog with well-measured reliability requires the ability to detect false positives and false alarms.  
Kepler effectively used automated vetting based on metrics that were tuned to remove detections from inverted Kepler data, and inverted data was used to measure Kepler reliability.  The observations in this paper lead us to expect that a similar strategy, appropriately tuned for TESS using inverted TESS data, will be successful. Automated vetting based, in part, on inverted light curves will be a significant step toward making TESS data useful for demographic studies.

\section{Acknowledgements}
We thank Michael Fausnaugh for useful insight into TESS systematics. This paper utilizes data from the Quick-Look Pipeline at the TESS Science Office at MIT. This work makes use of FFIs calibrated by TESS Image CAlibrator \citep{TICA}, which are also available as High-Level Science Products stored on the Mikulski Archive for Space Telescopes. Funding for the TESS mission is provided by NASA's Science Mission Directorate.

\bibliography{refs}

\newcommand{\noop}[1]{}
\begin{thebibliography}{11}
\expandafter\ifx\csname natexlab\endcsname\relax\def\natexlab#1{#1}\fi

\bibitem[{{Borucki} {et~al.}(2010){Borucki}, {Koch}, {Basri}, {Batalha},
  {Brown}, {Caldwell}, {Caldwell}, {Christensen-Dalsgaard}, {Cochran},
  {DeVore}, {Dunham}, {Dupree}, {Gautier}, {Geary}, {Gilliland}, {Gould},
  {Howell}, {Jenkins}, {Kondo}, {Latham}, {Marcy}, {Meibom}, {Kjeldsen},
  {Lissauer}, {Monet}, {Morrison}, {Sasselov}, {Tarter}, {Boss}, {Brownlee},
  {Owen}, {Buzasi}, {Charbonneau}, {Doyle}, {Fortney}, {Ford}, {Holman},
  {Seager}, {Steffen}, {Welsh}, {Rowe}, {Anderson}, {Buchhave}, {Ciardi},
  {Walkowicz}, {Sherry}, {Horch}, {Isaacson}, {Everett}, {Fischer}, {Torres},
  {Johnson}, {Endl}, {MacQueen}, {Bryson}, {Dotson}, {Haas}, {Kolodziejczak},
  {Van Cleve}, {Chandrasekaran}, {Twicken}, {Quintana}, {Clarke}, {Allen},
  {Li}, {Wu}, {Tenenbaum}, {Verner}, {Bruhweiler}, {Barnes}, \&
  {Prsa}}]{Borucki2010}
{Borucki}, W.~J., {Koch}, D., {Basri}, G., {Batalha}, N., {Brown}, T.,
  {Caldwell}, D., {Caldwell}, J., {Christensen-Dalsgaard}, J., {Cochran},
  W.~D., {DeVore}, E., {Dunham}, E.~W., {Dupree}, A.~K., {Gautier}, T.~N.,
  {Geary}, J.~C., {Gilliland}, R., {Gould}, A., {Howell}, S.~B., {Jenkins},
  J.~M., {Kondo}, Y., {Latham}, D.~W., {Marcy}, G.~W., {Meibom}, S.,
  {Kjeldsen}, H., {Lissauer}, J.~J., {Monet}, D.~G., {Morrison}, D.,
  {Sasselov}, D., {Tarter}, J., {Boss}, A., {Brownlee}, D., {Owen}, T.,
  {Buzasi}, D., {Charbonneau}, D., {Doyle}, L., {Fortney}, J., {Ford}, E.~B.,
  {Holman}, M.~J., {Seager}, S., {Steffen}, J.~H., {Welsh}, W.~F., {Rowe}, J.,
  {Anderson}, H., {Buchhave}, L., {Ciardi}, D., {Walkowicz}, L., {Sherry}, W.,
  {Horch}, E., {Isaacson}, H., {Everett}, M.~E., {Fischer}, D., {Torres}, G.,
  {Johnson}, J.~A., {Endl}, M., {MacQueen}, P., {Bryson}, S.~T., {Dotson}, J.,
  {Haas}, M., {Kolodziejczak}, J., {Van Cleve}, J., {Chandrasekaran}, H.,
  {Twicken}, J.~D., {Quintana}, E.~V., {Clarke}, B.~D., {Allen}, C., {Li}, J.,
  {Wu}, H., {Tenenbaum}, P., {Verner}, E., {Bruhweiler}, F., {Barnes}, J., \&
  {Prsa}, A. 2010, Science, 327, 977

\bibitem[{{Fausnaugh} {et~al.}(2020){Fausnaugh}, {Burke}, {Ricker}, \&
  {Vanderspek}}]{TICA}
{Fausnaugh}, M.~M., {Burke}, C.~J., {Ricker}, G.~R., \& {Vanderspek}, R. 2020,
  RNAAS, 4, 251

\bibitem[{{Feinstein} {et~al.}(2019){Feinstein}, {Montet}, {Foreman-Mackey},
  {Bedell}, {Saunders}, {Bean}, {Christiansen}, {Hedges}, {Luger}, {Scolnic},
  \& {Cardoso}}]{Feinstein2019}
{Feinstein}, A.~D., {Montet}, B.~T., {Foreman-Mackey}, D., {Bedell}, M.~E.,
  {Saunders}, N., {Bean}, J.~L., {Christiansen}, J.~L., {Hedges}, C., {Luger},
  R., {Scolnic}, D., \& {Cardoso}, J. V. d.~M. 2019, \pasp, 131, 094502

\bibitem[{{Hippke} {et~al.}(2019){Hippke}, {David}, {Mulders}, \&
  {Heller}}]{Hippke2019}
{Hippke}, M., {David}, T.~J., {Mulders}, G.~D., \& {Heller}, R. 2019, \aj, 158,
  143

\bibitem[{{Huang} {et~al.}(2020){Huang}, {Vanderburg}, {P{\'a}l}, {Sha}, {Yu},
  {Fong}, {Fausnaugh}, {Shporer}, {Guerrero}, {Vanderspek}, \&
  {Ricker}}]{Huang2020}
{Huang}, C.~X., {Vanderburg}, A., {P{\'a}l}, A., {Sha}, L., {Yu}, L., {Fong},
  W., {Fausnaugh}, M., {Shporer}, A., {Guerrero}, N., {Vanderspek}, R., \&
  {Ricker}, G. 2020, Research Notes of the American Astronomical Society, 4,
  204

\bibitem[{{Jenkins} {et~al.}(2016){Jenkins}, {Twicken}, {McCauliff},
  {Campbell}, {Sanderfer}, {Lung}, {Mansouri-Samani}, {Girouard}, {Tenenbaum},
  {Klaus}, {Smith}, {Caldwell}, {Chacon}, {Henze}, {Heiges}, {Latham},
  {Morgan}, {Swade}, {Rinehart}, \& {Vanderspek}}]{Jenkins2016}
{Jenkins}, J.~M., {Twicken}, J.~D., {McCauliff}, S., {Campbell}, J.,
  {Sanderfer}, D., {Lung}, D., {Mansouri-Samani}, M., {Girouard}, F.,
  {Tenenbaum}, P., {Klaus}, T., {Smith}, J.~C., {Caldwell}, D.~A., {Chacon},
  A.~D., {Henze}, C., {Heiges}, C., {Latham}, D.~W., {Morgan}, E., {Swade}, D.,
  {Rinehart}, S., \& {Vanderspek}, R. 2016, in Society of Photo-Optical
  Instrumentation Engineers (SPIE) Conference Series, Vol. 9913, Software and
  Cyberinfrastructure for Astronomy IV, ed. G.~{Chiozzi} \& J.~C. {Guzman},
  99133E

\bibitem[{{Kunimoto} {et~al.}(2021){Kunimoto}, {Huang}, {Tey}, {Fong}, {Hesse},
  {Shporer}, {Guerrero}, {Fausnaugh}, {Vanderspek}, \& {Ricker}}]{Kunimoto2021}
{Kunimoto}, M., {Huang}, C., {Tey}, E., {Fong}, W., {Hesse}, K., {Shporer}, A.,
  {Guerrero}, N., {Fausnaugh}, M., {Vanderspek}, R., \& {Ricker}, G. 2021,
  Research Notes of the American Astronomical Society, 5, 234

\bibitem[{{Kunimoto} {et~al.}(2022){Kunimoto}, {Tey}, {Fong}, {Hesse},
  {Shporer}, {Fausnaugh}, {Vanderspek}, \& {Ricker}}]{Kunimoto2022}
{Kunimoto}, M., {Tey}, E., {Fong}, W., {Hesse}, K., {Shporer}, A., {Fausnaugh},
  M., {Vanderspek}, R., \& {Ricker}, G. 2022, Research Notes of the American
  Astronomical Society, 6, 236

\bibitem[{{Paegert} {et~al.}(2021){Paegert}, {Stassun}, {Collins}, {Pepper},
  {Torres}, {Jenkins}, {Twicken}, \& {Latham}}]{Paegert2021}
{Paegert}, M., {Stassun}, K.~G., {Collins}, K.~A., {Pepper}, J., {Torres}, G.,
  {Jenkins}, J., {Twicken}, J.~D., \& {Latham}, D.~W. 2021, arXiv e-prints,
  arXiv:2108.04778

\bibitem[{{Ricker} {et~al.}(2015){Ricker}, {Winn}, {Vanderspek}, {Latham},
  {Bakos}, {Bean}, {Berta-Thompson}, {Brown}, {Buchhave}, {Butler}, {Butler},
  {Chaplin}, {Charbonneau}, {Christensen-Dalsgaard}, {Clampin}, {Deming},
  {Doty}, {De Lee}, {Dressing}, {Dunham}, {Endl}, {Fressin}, {Ge}, {Henning},
  {Holman}, {Howard}, {Ida}, {Jenkins}, {Jernigan}, {Johnson}, {Kaltenegger},
  {Kawai}, {Kjeldsen}, {Laughlin}, {Levine}, {Lin}, {Lissauer}, {MacQueen},
  {Marcy}, {McCullough}, {Morton}, {Narita}, {Paegert}, {Palle}, {Pepe},
  {Pepper}, {Quirrenbach}, {Rinehart}, {Sasselov}, {Sato}, {Seager},
  {Sozzetti}, {Stassun}, {Sullivan}, {Szentgyorgyi}, {Torres}, {Udry}, \&
  {Villasenor}}]{Ricker2015}
{Ricker}, G.~R., {Winn}, J.~N., {Vanderspek}, R., {Latham}, D.~W., {Bakos},
  G.~{\'A}., {Bean}, J.~L., {Berta-Thompson}, Z.~K., {Brown}, T.~M.,
  {Buchhave}, L., {Butler}, N.~R., {Butler}, R.~P., {Chaplin}, W.~J.,
  {Charbonneau}, D., {Christensen-Dalsgaard}, J., {Clampin}, M., {Deming}, D.,
  {Doty}, J., {De Lee}, N., {Dressing}, C., {Dunham}, E.~W., {Endl}, M.,
  {Fressin}, F., {Ge}, J., {Henning}, T., {Holman}, M.~J., {Howard}, A.~W.,
  {Ida}, S., {Jenkins}, J.~M., {Jernigan}, G., {Johnson}, J.~A., {Kaltenegger},
  L., {Kawai}, N., {Kjeldsen}, H., {Laughlin}, G., {Levine}, A.~M., {Lin}, D.,
  {Lissauer}, J.~J., {MacQueen}, P., {Marcy}, G., {McCullough}, P.~R.,
  {Morton}, T.~D., {Narita}, N., {Paegert}, M., {Palle}, E., {Pepe}, F.,
  {Pepper}, J., {Quirrenbach}, A., {Rinehart}, S.~A., {Sasselov}, D., {Sato},
  B., {Seager}, S., {Sozzetti}, A., {Stassun}, K.~G., {Sullivan}, P.,
  {Szentgyorgyi}, A., {Torres}, G., {Udry}, S., \& {Villasenor}, J. 2015,
  Journal of Astronomical Telescopes, Instruments, and Systems, 1, 014003

\bibitem[{{Thompson} {et~al.}(2018){Thompson}, {Coughlin}, {Hoffman},
  {Mullally}, {Christiansen}, {Burke}, {Bryson}, {Batalha}, {Haas},
  {Catanzarite}, {Rowe}, {Barentsen}, {Caldwell}, {Clarke}, {Jenkins}, {Li},
  {Latham}, {Lissauer}, {Mathur}, {Morris}, {Seader}, {Smith}, {Klaus},
  {Twicken}, {Van Cleve}, {Wohler}, {Akeson}, {Ciardi}, {Cochran}, {Henze},
  {Howell}, {Huber}, {Pr{\v s}a}, {Ram{\'{\i}}rez}, {Morton}, {Barclay},
  {Campbell}, {Chaplin}, {Charbonneau}, {Christensen-Dalsgaard}, {Dotson},
  {Doyle}, {Dunham}, {Dupree}, {Ford}, {Geary}, {Girouard}, {Isaacson},
  {Kjeldsen}, {Quintana}, {Ragozzine}, {Shabram}, {Shporer}, {Silva Aguirre},
  {Steffen}, {Still}, {Tenenbaum}, {Welsh}, {Wolfgang}, {Zamudio}, {Koch}, \&
  {Borucki}}]{Thompson2018}
{Thompson}, S.~E., {Coughlin}, J.~L., {Hoffman}, K., {Mullally}, F.,
  {Christiansen}, J.~L., {Burke}, C.~J., {Bryson}, S., {Batalha}, N., {Haas},
  M.~R., {Catanzarite}, J., {Rowe}, J.~F., {Barentsen}, G., {Caldwell}, D.~A.,
  {Clarke}, B.~D., {Jenkins}, J.~M., {Li}, J., {Latham}, D.~W., {Lissauer},
  J.~J., {Mathur}, S., {Morris}, R.~L., {Seader}, S.~E., {Smith}, J.~C.,
  {Klaus}, T.~C., {Twicken}, J.~D., {Van Cleve}, J.~E., {Wohler}, B., {Akeson},
  R., {Ciardi}, D.~R., {Cochran}, W.~D., {Henze}, C.~E., {Howell}, S.~B.,
  {Huber}, D., {Pr{\v s}a}, A., {Ram{\'{\i}}rez}, S.~V., {Morton}, T.~D.,
  {Barclay}, T., {Campbell}, J.~R., {Chaplin}, W.~J., {Charbonneau}, D.,
  {Christensen-Dalsgaard}, J., {Dotson}, J.~L., {Doyle}, L., {Dunham}, E.~W.,
  {Dupree}, A.~K., {Ford}, E.~B., {Geary}, J.~C., {Girouard}, F.~R.,
  {Isaacson}, H., {Kjeldsen}, H., {Quintana}, E.~V., {Ragozzine}, D.,
  {Shabram}, M., {Shporer}, A., {Silva Aguirre}, V., {Steffen}, J.~H., {Still},
  M., {Tenenbaum}, P., {Welsh}, W.~F., {Wolfgang}, A., {Zamudio}, K.~A.,
  {Koch}, D.~G., \& {Borucki}, W.~J. 2018, Astrophysical Journal Supplement,
  235, 38

\end{thebibliography}


\end{document}